\documentclass[aps,prb,reprint,showpacs]{revtex4-1}
\usepackage{graphicx}
\begin{document}

\title{Structural position and oxidation state of nickel in SrTiO$_3$}

\author{Irina A. Sluchinskaya}
\email[]{irinasluch@nm.ru}
\affiliation{Physics Department, Moscow State University, \\
Leninskie gory, 119991 Moscow, Russia}
\author{Alexander I. Lebedev}
%\email[]{swan@scon155.phys.msu.ru}
\affiliation{Physics Department, Moscow State University, \\
Leninskie gory, 119991 Moscow, Russia}
\author{Alexei Erko}
\affiliation{Helmholtz-Zentrum Berlin, \\
Albert-Einstein-Str. 15, 12489 Berlin, Germany}

\date{\today}

\begin{abstract}
The properties of Ni-doped strontium titanate are studied using X-ray
diffraction and XAFS spectroscopy. It is shown that regardless of the preparation
conditions, the SrTi$_{1-x}$Ni$_x$O$_3$ solid solution and the NiTiO$_3$ phase
are the most stable phases which can coexist. According to the EXAFS data, in
the single-phase sample of SrTi$_{0.97}$Ni$_{0.03}$O$_3$, the Ni atoms substitute
for the Ti ones and are on-center. The distortion of the oxygen octahedra
is not observed. The XANES spectra analysis shows that the oxidation
state of nickel in NiTiO$_3$ is 2+, and in the SrTi$_{1-x}$Ni$_x$O$_3$ solid
solution it is close to 4+. It is shown that the strongest light absorption in
doped samples is associated with the presence of tetravalent nickel in the
SrTi$_{1-x}$Ni$_x$O$_3$ solid solution. This doping seems the most promising one
for solar energy converters that exploit the bulk photovoltaic effect.
\end{abstract}

\pacs{61.10.Ht, 72.50.Pz, 77.84.Dy, 84.60.Jt}

\maketitle

\section{Introduction}

The bulk photovoltaic effect consists in the occurrence of a photocurrent
and very large photovoltages when illuminating homogeneous crystals having no
inversion symmetry.~\cite{Sturman-Fridkin}  The idea of the practical application
of this effect in ferroelectrics for solar energy conversion was discussed
back in the 70s.~\cite{ApplPhysLett.25.233}  However, because of the short
lifetime of photo-excited carriers, the quantum yield of this effect is
generally small, and the idea was decided to be unproductive. In recent years,
the interest to the ferroelectric oxides with the perovskite structure has
revived because new ideas how to increase the efficiency of solar energy
converters based on the bulk photovoltaic effect have been
proposed.~\cite{ApplPhysLett.93.122904,NatureCommun.2.256}
The main disadvantage of the ferroelectric oxides is their relatively large
band gap, causing them to absorb only a small fraction of the solar radiation.
Recent theoretical studies have shown that the substitution of Ti atoms at
the $B$~sites of the perovskite structure with divalent impurities having
the $d^8$ electron configuration (Ni, Pd, Pt), compensated by the oxygen
vacancy, decreases the band gap and the obtained perovskites are polar
semiconductor oxides.~\cite{JAmChemSoc.130.17409,PhysRevB.83.205115}

An additional interest to the study of the Ni impurity is associated with
the results obtained from the recent experimental and theoretical studies
of new materials---a recently synthesized PbNiO$_3$ which has a very high
calculated spontaneous
polarization~\cite{JPhysConfSer.215.012131,JAmChemSoc.133.16920,PhysRevB.86.014116}
and BiNiO$_3$ with unexpected oxidation states of nickel and bismuth
atoms.~\cite{JAmChemSoc.129.14433}

In addition, the search for new magnetic off-center impurities in incipient
ferroelectrics is still important because they can result in simultaneous
emergence of the ferroelectricity and magnetic ordering and give rise to the
magnetoelectric interaction. Materials with these properties belong to
multiferroics. SrTiO$_3$ doped with Mn at the $A$ site is an example of such
a material, in which a new type of magnetoelectric interaction was recently
discovered.~\cite{PhysRevLett.101.165704,EurPhysJB.71.407}  Ni-doped samples
could be an another example.

Since the doping impurity can enter several different sites in the perovskite
structure and stay in them in different oxidation states, the aim of this work
was to study the structural position and the oxidation state of the Ni impurity
in SrTiO$_3$ prepared under different conditions using XAFS spectroscopy. We
planned to check the possibility of preparing samples doped with divalent Ni
at the $B$~site, to evaluate the possibility of incorporating the Ni impurity
into the $A$~site, and to establish a correlation between the optical properties
of the samples, on the one hand, and the structural position and the oxidation
state of the Ni impurity, on the other hand. The choice of SrTiO$_3$ was
dictated by the fact that earlier we have studied the structural position and
the oxidation state of a number of $3d$ elements in SrTiO$_3$
(Mn,~\cite{JETPLett.89.457,BullRASPhys.74.1235}  Co,~\cite{VKS19.116}
Fe~\cite{RSNE8.347}), and their combined analysis can allow to find new promising
impurities for solar energy converters.

\section{Samples and experimental techniques}

Ni-doped samples of SrTiO$_3$ with the impurity concentration of 2--3\% and
different deviation from stoichiometry were prepared by the solid-state
reaction method. The starting materials were SrCO$_3$, nanocrystalline TiO$_2$,
and Ni(CH$_3$COO)$_2$$\cdot$4H$_2$O. The components were weighed in proper
proportions, mixed, ground in acetone, and calcined in air at 1100$^\circ$C
for 8~h. The calcined powders were ground once more and annealed again under
the same conditions. Some of the samples were additionally calcined in air
at 1500$^\circ$C for 2~h. The composition of the samples was deliberately
deviated from stoichiometry (excess titanium or excess strontium) in
order to incorporate the impurity into the $A$ or $B$~site of the perovskite
structure.

The reference compounds NiO, NiTiO$_3$, and BaNiO$_{3-\delta}$ used for
determination of the oxidation state of Ni in SrTiO$_3$ were prepared as
follows. NiO was obtained by thermal decomposition of
Ni(CH$_3$COO)$_2$$\cdot$4H$_2$O. Two other samples were prepared by the
solid-state reaction method in air: the NiTiO$_3$ sample was obtained from
Ni(CH$_3$COO)$_2$$\cdot$4H$_2$O and TiO$_2$ at 1100$^\circ$C, the
BaNiO$_{3-\delta}$ sample was prepared from BaO$_2$ and NiO at 650$^\circ$C.
The phase composition of the samples was controlled by X-ray diffraction.

Extended X-ray absorption fine structure (EXAFS) and X-ray absorption near-edge
structure (XANES) spectra were obtained at KMC-2 station of the BESSY
synchrotron radiation source (the beam energy 1.7 GeV; the beam current up to
290 mA) at the Ni $K$-edge (8340~eV) at 300~K. The radiation was monochromatized
by a double-crystal Si$_{1-x}$Ge$_x$(111) monochromator. Spectra were collected
in fluorescence mode. The radiation intensity incident on the sample ($I_0$)
was measured by an ionization chamber; the fluorescence intensity ($I_f$) was
measured by a silicon energy-dispersive R\"ONTEC X-flash detector with
10~mm$^2$ active area.

Isolation of the oscillating EXAFS function $\chi(k)$ from the fluorescence
excitation spectra $\mu (E) = I_f/I_0$ (where $E$ is the X-ray photon energy)
was performed in the traditional way.~\cite{IzvAkadNaukSerFiz.60.46,PhysRevB.55.14770}
After subtracting the pre-edge background, splines were used to extract the monotonic
atomic part of the spectrum $\mu_0 (E)$ and then the dependence of
$\chi = (\mu - \mu_0) / \mu_0$ was calculated as a function of the photoelectron
wave vector $k = (2m (E - E_0)/\hbar^2)^{1/2}$. The energy origin, $E_0$, was
taken to be the position of the inflection point on the absorption edge. For each
sample three spectra were recorded, they were then independently processed, and
the resulting $\chi(k)$ curves finally averaged.

Direct and inverse Fourier transforms with modified Hanning windows were used
to extract the information about the first three shells from the obtained $\chi(k)$
curves. The distances $R_j$ and Debye-Waller factors $\sigma^2_j$ for $j$th
shell ($j = 1$--3) as well as the energy origin correction $dE_0$ were
simultaneously varied to obtain the minimum root-mean-square deviation between
the experimental and calculated $k^2 \chi(k)$ curves. The coordination numbers
were considered fixed for a given structural model. The number of adjustable
parameters (8) was usually about a half of the number of independent data
points $N_{\rm ind} = 2\Delta k \Delta R/\pi \approx 16$.

The single- and multiple-scattering amplitudes and phase shifts, the central
atom phase shift, and the photoelectron mean free path as a function of $k$,
needed to calculate the theoretical curves $\chi(k)$, were computed using the
\texttt{FEFF6} software.~\cite{FEFF}

EXAFS spectra were also analyzed with a widely used \texttt{IFEFFIT} software
package.~\cite{IFEFFIT}  Isolation of the experimental EXAFS function was
carried out using the ATHENA program and its fitting to the theoretical curve
calculated for a given structural model was performed using the ARTEMIS program.
In this approach, the amplitudes and phase shifts for all single- and
multiple-scattering paths were also calculated using the \texttt{FEFF6} software. The
results obtained by two different data analysis approaches agreed well.

\section{Results}
\subsection{X-ray data}

\begin{figure}
\includegraphics{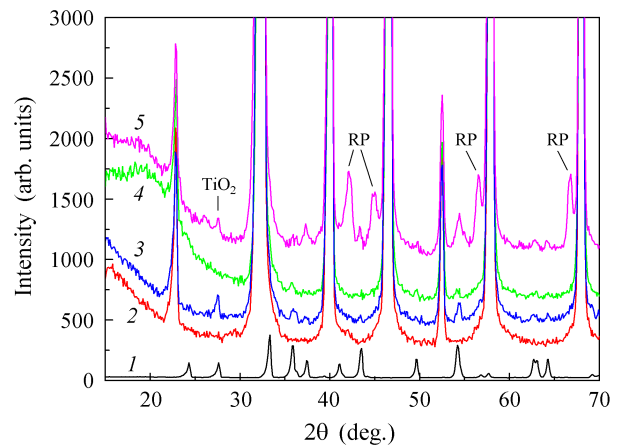}
\caption{\label{fig1}(Color online) Diffraction patterns of samples: (1)~NiTiO$_3$,
(2)~SrTi$_{0.97}$Ni$_{0.03}$O$_3$ sample annealed at 1500$^\circ$C,
(3)~Sr$_{0.98}$Ni$_{0.02}$TiO$_3$ sample annealed at 1100$^\circ$C,
(4)~Sr$_{0.98}$Ni$_{0.02}$TiO$_3$ sample annealed at 1500$^\circ$C,
(5)~SrTi$_{0.97}$Ni$_{0.03}$O$_3$ sample annealed at 1100$^\circ$C.
Reflections of TiO$_2$ and Ruddlesden--Popper (RP) phases are also indicated.}
\end{figure}

The diffraction patterns of all investigated samples are shown in Fig.~\ref{fig1}.
It is seen that the SrTi$_{0.97}$Ni$_{0.03}$O$_3$ sample annealed at 1500$^\circ$C
is the only single-phase sample which has a cubic perovskite structure; for
other samples, additional reflections were observed on the diffraction patterns.
For the SrTi$_{0.97}$Ni$_{0.03}$O$_3$
and Sr$_{0.98}$Ni$_{0.02}$TiO$_3$ samples annealed at 1100$^\circ$C, along with
the reflections characteristic of the perovskite phase, additional reflections
indicating a small amount of TiO$_2$ and, presumably, NiTiO$_3$
were observed. The identification of a possible NiO phase was complicated by the
closeness of its reflections to the position of NiTiO$_3$ reflections. In addition,
stronger lines of the Ruddlesden--Popper phases Sr$_3$Ti$_2$O$_7$ and
Sr$_4$Ti$_3$O$_{10}$ were observed in the SrTi$_{0.97}$Ni$_{0.03}$O$_3$ sample
annealed at 1100$^\circ$C. In the Sr$_{0.98}$Ni$_{0.02}$TiO$_3$ sample annealed at
1500$^\circ$C the only additional phase, NiTiO$_3$, was found.

Since barium nickelate BaNiO$_{3-\delta}$ is a defective phase with the $\delta$
value depending on the preparation conditions, its lattice parameters dependence
on the $\delta$ value~\cite{JChemSoc:DaltonTrans.1975.1055} was used to determine
the actual oxygen content in our BaNiO$_{3-\delta}$ sample. The hexagonal lattice
parameters of $a = 5.568 \pm 0.001$~{\AA}, $c = 4.838 \pm 0.001$~{\AA} in our
sample correspond to $\delta \approx 0.4$.

\subsection{XANES data analysis}

\begin{figure}
\includegraphics{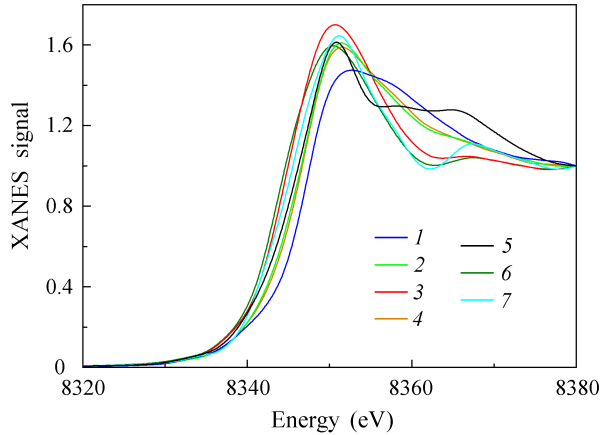}
\caption{\label{fig2}(Color online) XANES spectra for SrTiO$_3$(Ni) samples
and nickel reference compounds:
(1)~SrTi$_{0.97}$Ni$_{0.03}$O$_3$ sample annealed at 1500$^\circ$C,
(2)~SrTi$_{0.97}$Ni$_{0.03}$O$_3$ sample annealed at 1100$^\circ$C,
(3)~Sr$_{0.98}$Ni$_{0.02}$TiO$_3$ sample annealed at 1500$^\circ$C,
(4)~Sr$_{0.98}$Ni$_{0.02}$TiO$_3$ sample annealed at 1100$^\circ$C,
(5)~BaNiO$_{3-\delta}$, (6)~NiTiO$_3$, (7)~NiO.}
\end{figure}

To determine the oxidation state of the Ni impurity in SrTiO$_3$, the position
of the absorption edge in XANES spectra of the samples was compared with the
edge positions in the reference compounds. XANES spectra of all studied samples
and three reference compounds are shown in Fig.~\ref{fig2}.

Comparison of the spectra of the Sr$_{0.98}$Ni$_{0.02}$TiO$_3$ sample annealed
at 1500$^\circ$C with the spectra of cubic NiO and rhombohedral NiTiO$_3$ (with
the ilmenite structure) shows that the absorption edges in these samples are
very close. So, we conclude that the Ni impurity in the sample under
consideration is in the 2+ oxidation
state. The absorption edges of the SrTi$_{0.97}$Ni$_{0.03}$O$_3$ and
Sr$_{0.98}$Ni$_{0.02}$TiO$_3$ samples annealed at 1100$^\circ$C are close to the
absorption edge of the BaNiO$_{3-\delta}$ reference compound.
In the single-phase SrTi$_{0.97}$Ni$_{0.03}$O$_3$
sample annealed at 1500$^\circ$C, the absorption edge is shifted to even higher
energies (by 2.5~eV as compared to NiO, by 2.9~eV as compared to NiTiO$_3$,
and by 1.3~eV as compared to BaNiO$_{3-\delta}$).

\subsection{EXAFS data analysis}

\begin{figure}
\includegraphics{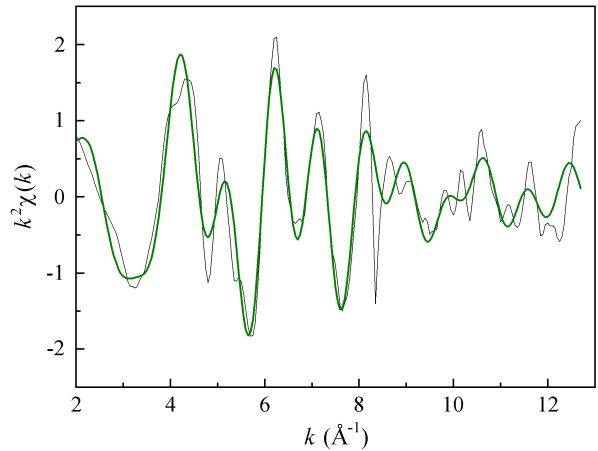}
\caption{\label{fig3}(Color online) EXAFS spectrum recorded at the Ni $K$-edge
at 300~K for SrTi$_{0.97}$Ni$_{0.03}$O$_3$ sample annealed at 1500$^\circ$C
(thin black line) and its best theoretical fit (thick green line).}
\end{figure}

\begin{table}
\caption{\label{table1}Structural parameters obtained from the data analysis
of the EXAFS spectra for SrTi$_{0.97}$Ni$_{0.03}$O$_3$ sample annealed at
1500$^\circ$C ($R_i$ is the distance to the $i$th shell, $\sigma^2_i$ is the
Debye-Waller factor for this shell).}
\begin{ruledtabular}
\begin{tabular}{cccc}
Shell & $R_i$~(\AA) & $\sigma_i^2$~(\AA$^2$) & Atom \\
\hline
1     & 1.914$\pm$0.004 & 0.0035$\pm$0.0006  & O \\
2     & 3.342$\pm$0.006 & 0.0084$\pm$0.0007  & Sr \\
3     & 3.877$\pm$0.004 & 0.0053$\pm$0.0005  & Ti \\
\end{tabular}
\end{ruledtabular}
\end{table}

To determine the structural position of the Ni impurity, EXAFS spectra were
analyzed. A typical EXAFS spectrum $k^2 \chi(k)$ for the single-phase
SrTi$_{0.97}$Ni$_{0.03}$O$_3$ sample annealed at 1500$^\circ$C and its best
theoretical fit (which takes into account the multiple-scattering effects)
are shown in Fig.~\ref{fig3}. The best agreement between the calculated and
experimental data was obtained in the model in which Ni atoms substitute for
Ti atoms in SrTiO$_3$. The interatomic distances and Debye-Waller factors
for three nearest shells are given in Table~\ref{table1}. Small values of
the Debye-Waller factor for the first and second shells, which are typical
for the thermal vibrations in perovskites at 300~K, enable to draw two
conclusions: (1)~the off-centering of Ni atoms at the $B$~sites can be
excluded and (2)~there is no distortion of the oxygen octahedra around
the impurity atoms.%
    \footnote{The sensitivity of the Debye-Waller factor to small distortions
    of the oxygen octahedra can be illustrated by the EXAFS data for NiTiO$_3$.
    In this compound, there are two Ni-O distances in the first shell which
    differ by $\sim$0.07~{\AA}; such a small distortion increases the Debye-Waller
    factor to 0.0073$\pm$0.0012~{\AA}$^2$.}

\begin{figure}
\includegraphics{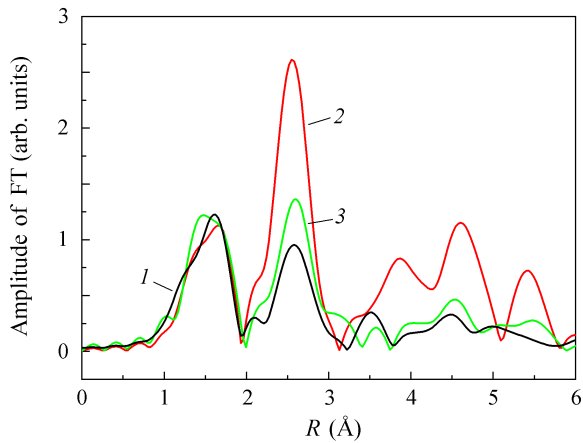}
\caption{\label{fig4}(Color online) Comparison of Fourier transforms of the
EXAFS $k^2 \chi(k)$ spectra obtained for (1)~Sr$_{0.98}$Ni$_{0.02}$TiO$_3$
sample annealed at 1500$^\circ$C and for (2)~NiO and (3)~NiTiO$_3$ reference
compounds.}
\end{figure}

The analysis of the EXAFS spectra helped us to determine the composition of
the second phase precipitating in the Sr$_{0.98}$Ni$_{0.02}$TiO$_3$ sample
annealed at 1500$^\circ$C. Although the distances and
the coordination numbers for the first shell of Ni are close in NiO and NiTiO$_3$,
the coordination numbers for the second shell in these compounds differ by
three times. A comparison of the Fourier transforms of the EXAFS spectra for
the sample under consideration and two reference compounds, NiTiO$_3$ and NiO,
shows a better agreement of its spectrum with that of NiTiO$_3$ (Fig.~\ref{fig4}).
This means that among two possible phases, NiTiO$_3$ and NiO, the second
phase in our samples is the NiTiO$_3$ one.

As concerns to two-phase SrTi$_{0.97}$Ni$_{0.03}$O$_3$ and
Sr$_{0.98}$Ni$_{0.02}$TiO$_3$ samples annealed at 1100$^\circ$C, the comparison
of their EXAFS spectra with the EXAFS spectrum of the single-phase solid
solution and the EXAFS spectrum of NiTiO$_3$ shows that the spectra of the
samples under discussion may be represented as a superposition of the spectra
of NiTiO$_3$ and of the solid solution in a ratio close to 1:1.

The optical properties of the samples are consistent with the data obtained
above. The Sr$_{0.98}$Ni$_{0.02}$TiO$_3$ sample annealed at 1500$^\circ$C had a
light brown color; the SrTi$_{0.97}$Ni$_{0.03}$O$_3$ and Sr$_{0.98}$Ni$_{0.02}$TiO$_3$
samples annealed at 1100$^\circ$C were dark brown; the SrTi$_{0.97}$Ni$_{0.03}$O$_3$
sample annealed at 1500$^\circ$C had an almost black color. Thus, the color of
the samples reflects the relative amounts of black SrTi$_{1-x}$Ni$_x$O$_3$ and
yellow NiTiO$_3$ phases in the samples.

\section{Discussion}

Combined analysis of X-ray and EXAFS data shows that the SrTi$_{0.97}$Ni$_{0.03}$O$_3$
sample annealed at 1500$^\circ$C is the single-phase solid solution in which
the Ni atoms substitute for the Ti atoms at the $B$~sites and are on-center.
This means that the solubility of nickel at the $B$~sites in SrTiO$_3$ exceeds
3\% at 1500$^\circ$C. In the sample with a nominal composition of
SrTi$_{0.97}$Ni$_{0.03}$O$_3$ annealed at 1100$^\circ$C, the appearance of
reflections of NiTiO$_3$ and of the Ruddlesden-Popper phases indicates that
a part of Ni atoms is spent on the NiTiO$_3$ formation, while the other part
remains at the $B$~sites. As a result, some untapped Sr atoms form the SrO
planes which embed into the perovskite structure and form the
Sr$_3$Ti$_2$O$_7$ and Sr$_4$Ti$_3$O$_{10}$ phases.

When trying to incorporate Ni atoms into the $A$~sites of SrTiO$_3$ at
1500$^\circ$C (the samples with a nominal composition of
Sr$_{0.98}$Ni$_{0.02}$TiO$_3$), the precipitation of the NiTiO$_3$ second phase
occurs. The concentration of nickel in the solid solution phase is small, as it
follows from the EXAFS data and the sample color. In contrast,
when annealing the Sr$_{0.98}$Ni$_{0.02}$TiO$_3$ sample at a lower temperature
(1100$^\circ$C), nickel in the sample is in a mixture of NiTiO$_3$ and
SrTi$_{1-x}$Ni$_x$O$_3$ solid solution as follows from the XANES data, EXAFS
data, and the sample color. The precipitation of a small portion of TiO$_2$
confirms the incorporation of Ni into SrTiO$_3$ (the appearance of TiO$_2$ is
a consequence of the removing of some Ti atoms from the $B$~sites when doping
strontium titanate with nickel).

As concerns to the appearance of the TiO$_2$ phase in the SrTi$_{0.97}$Ni$_{0.03}$O$_3$
sample annealed at 1100$^\circ$C, we suppose that its appearance is possible
for kinetic reasons. The formation of the solid solution occurs as a result of
a chain of chemical reactions in which the NiTiO$_3$ phase initially formed at
$\sim$750$^\circ$C from NiO and TiO$_2$ reacts with excess SrO resulted from
the thermal decomposition of SrCO$_3$ at $\sim$1000$^\circ$C to produce
SrNiO$_{2.5}$+TiO$_2$ as intermediate phases. This reaction is quite slow
(the preparation of SrNiO$_{2.5}$ usually takes 48~h
at 1000$^\circ$C~\cite{JInorgNuclChem.34.1599}). As a result, in the samples
annealed at 1100$^\circ$C for 16 h,
a mixture of the solid solution, NiTiO$_3$, and TiO$_2$ is observed. At higher
annealing temperature, the kinetic processes are faster, the reaction completes,
and the composition of the samples is fully controlled by the deviation from
stoichiometry.

Therefore, the stable phases in the samples are the single-phase SrTi$_{1-x}$Ni$_x$O$_3$
solid solution and NiTiO$_3$; their ratio depends on the deviation from
stoichiometry and the annealing temperature. Nickel cannot be incorporated into
the $A$~sites of strontium titanate. The low stability of Ni at the $A$~site of
SrTiO$_3$ is apparently related to a large difference in the ionic radii of
Ni$^{2+}$ and Sr$^{2+}$. The 12-fold coordination is not typical for nickel;
for the coordination number of 6, the ionic radius of Ni$^{2+}$ (0.69~{\AA})
is much less than that of Sr$^{2+}$ (1.18~{\AA}).~\cite{ActaCrystA.32.751}

The XANES data, which were used to determine the oxidation state of nickel, are
fully consistent with X-ray and EXAFS data. In the sample with a nominal
composition of Sr$_{0.98}$Ni$_{0.02}$TiO$_3$
annealed at 1500$^\circ$C, in which nickel is in the NiTiO$_3$ phase, the
oxidation state of Ni coincides with its oxidation state in NiO and NiTiO$_3$ and
is 2+. In the single-phase SrTi$_{0.97}$Ni$_{0.03}$O$_3$ sample annealed at
1500$^\circ$C, the shift of the absorption edge with respect to NiTiO$_3$ is
maximum and is about twice the shift between the reference compounds NiTiO$_3$
and BaNiO$_{3-\delta}$. If, following Ref.~\onlinecite{JChemSoc:DaltonTrans.1975.1055},
we start from the number of ions, their nominal charge, and the
$\delta \approx 0.4$ value determined from the lattice parameters,
the average oxidation state of nickel in BaNiO$_{3-\delta}$ is
$(4-2\delta) \approx 3.2$. Then, according to the shift of the absorption edge in
SrTi$_{0.97}$Ni$_{0.03}$O$_3$ sample annealed at 1500$^\circ$C, the oxidation
state of nickel in it should be close to 4+. For the SrTi$_{0.97}$Ni$_{0.03}$O$_3$
and Sr$_{0.98}$Ni$_{0.02}$TiO$_3$ samples annealed at 1100$^\circ$C, which are
the mixture of two nickel-containing phases in proportion close to 1:1,
the position of the
absorption edge is intermediate between those in two stable phases and is close
to the position of the absorption edge in BaNiO$_{3-\delta}$ (about 3+).

It should be noted that the question about the oxidation state of Ni in
SrTi$_{0.97}$Ni$_{0.03}$O$_3$ is not so simple. The discussion about the
oxidation state of Ni in BaNiO$_{3-\delta}$ is still going on, in particular,
the doubts were expressed~\cite{InorgChem.37.1513} about the validity of its
formal determination based on the number of ions and their nominal charges.
In the cited work, the data of M\"ossbauer spectroscopy for BaNiO$_3$ indicated
the Ni oxidation state close to 4+, whereas the photoelectron spectroscopy gave
a value close to 3+. The authors of Ref.~\onlinecite{InorgChem.37.1513} suggested
a model in which Ni is trivalent and its charge is compensated by a hole bound
by one or two negative oxygen ions. Within this model, the Ni oxidation state
determined from the absorption edge shift in our SrTi$_{0.97}$Ni$_{0.03}$O$_3$
solid solution should be closer to 3+. This explanation, however, contradicts
the fact that our EXAFS data analysis revealed no distortion of the oxygen
octahedra, whereas the localization of a hole at one or two oxygen ions should
cause its distortion.

At the same time, the experimental study of XANES spectra of Li$_x$NiO$_2$
compound used in lithium batteries have shown that a variation in the degree
of intercalation $x$ changes the Ni oxidation state from 2+ to 4+ and shifts
the absorption edge in XANES spectra by $\sim$3.5~eV.~\cite{JPhysChemA.102.65}
This shift is close to the shift of 2.9~eV observed in our spectra
between NiTiO$_3$ and the SrTi$_{0.97}$Ni$_{0.03}$O$_3$ sample annealed at
1500$^\circ$C. Moreover, the interatomic Ni-O distance (1.914~{\AA}) obtained
from our EXAFS measurements is less than the sum of the ionic radii of
Ni$^{3+}$ and O$^{2-}$ (0.56~+~1.4 = 1.96~{\AA}) and is closer to the sum of
the ionic radii of Ni$^{4+}$ and O$^{2-}$ (0.48~+~1.4 = 1.88~{\AA}). Yet
another argument in favor of tetravalent nickel can be the small value of the
Debye-Waller factor $\sigma_1^2$, which indicates the absence of the Ni
displacement from the $B$~site in the SrTi$_{0.97}$Ni$_{0.03}$O$_3$ solid
solution. If the nickel atom in this phase was in the Ni$^{3+}$ oxidation state,
it required to be charge-compensated by the oxygen vacancy located nearby the
impurity atom (such Ni$^{3+}$-$V_{\rm O}$ axial centers with the Ni displacement
from the center of the octahedron up to $\sim$0.3~{\AA} have been observed in
electron spin resonance (ESR) spectra~\cite{PhysRev.186.361,PhysScr.75.147}).
However, our EXAFS measurements did not reveal noticeable distortion of the
oxygen octahedra surrounding the Ni atom.

It should be noted that the EXAFS data analysis of the SrTi$_{0.97}$Ni$_{0.03}$O$_3$
sample annealed at 1500$^\circ$C revealed a reduced value of 4.90$\pm$0.34 for
the coordination number for the first shell of Ni. There are two possible
explanation of this fact. First, it can be an evidence for the existence
of a small amount of Ni$^{3+}$-$V_{\rm O}$ complexes. The coordination number
for these complexes is effectively reduced to 4 because one of the oxygen atoms
is missing and the other one is located at a different distance compared to
four remaining O atoms and does not contribute much to the EXAFS signal. This
explanation agrees with the reduced shift of the absorption edge in our sample
with respect to the Ni$^{4+}$ state in Li$_x$NiO$_2$. Another explanation is
the existence of a small amount of the NiTiO$_3$ phase located at the grain
boundaries; the experiment shows that the contamination with this phase strongly
decreases the coordination number because of out-of-phase EXAFS oscillations
in two nickel-containing phases.

Neverthless, we think that the oxidation state of Ni at the $B$~site in
SrTiO$_3$ is close to 4+; this disagrees with the suggestion made
earlier~\cite{JAmChemSoc.130.17409,PhysRevB.83.205115} that Ni is in the 2+
oxidation state in the related ferroelectric PbTiO$_3$. Although the
question about the actual oxidation state of Ni in the latter material should
be tested experimentally, our findings indicate that the strong light
absorption in Ni-doped SrTiO$_3$ is associated with the tetravalent nickel.

From the viewpoint of possible application of doped perovskites in solar
energy conversion it is interesting to compare the properties of
SrTiO$_3$ doped with nickel and with other
$3d$ elements.~\cite{JETPLett.89.457,BullRASPhys.74.1235,VKS19.116,RSNE8.347}
The absorption spectra of doped samples are systematically shifted to the
infrared region with increasing atomic number from Mn to Ni: manganese-doped
samples are greenish brown, iron-doped samples are brown, cobalt-doped samples
are dark brown, and Ni gives an almost black color to the samples. Thus,
for creating samples that strongly absorb light in the whole visible region,
the nickel doping seems the most promising. Intense absorption in the samples
suggests that it is associated with charge-transfer transitions.

Interestingly, the small value of the Debye-Waller factor for the first shell
($\sigma_1^2 \approx {}$0.0035~{\AA}$^2$) also excludes the possibility of
Jahn-Teller instability of the Ni$^{4+}$ ion in SrTiO$_3$, which is possible
for the octahedral $d^6$ configuration. In Li$_x$NiO$_2$, the Jahn-Teller
instability of Ni$^{4+}$ manifests itself in the distorted oxygen octahedra
with Ni-O bond lengths of 1.88 and 2.08~{\AA}. In the EXAFS spectra, this
distorted environment should be observed as one O shell with an average
interatomic distance of 1.947~{\AA} and a static Debye-Waller factor of
0.009~{\AA}$^2$. The experimental value of $\sigma_1^2$ in SrTiO$_3$(Ni) is
much lower than the value estimated for Jahn-Teller-distorted environment.

The interpretation of our data on the oxidation state of nickel in SrTiO$_3$
differs much from the results of earlier studies. The comparison with the
data of Ref.~\onlinecite{JMaterChem.19.4391}, in which SrTiO$_3$(Ni) samples
were also studied by XAFS technique, shows that the XANES and EXAFS spectra
obtained in our work and in Ref.~\onlinecite{JMaterChem.19.4391} are qualitatively
different. For example, in Ref.~\onlinecite{JMaterChem.19.4391} the shift of
the absorption edge in SrTiO$_3$(Ni) with respect to NiO was only 1.1~eV
(in our samples it is 2.5~eV). Moreover, even the color of the samples is
different (beige in Ref.~\onlinecite{JMaterChem.19.4391} and almost black in
our work). We believe that these differences in the properties of the samples
result from the different methods of their preparation (solid-state
reaction method in our case and hydrothermal synthesis at 150$^\circ$C in
Ref.~\onlinecite{JMaterChem.19.4391}). In
Ref.~\onlinecite{PhysRev.186.361,PhysScr.75.147}, single crystals of
SrTiO$_3$(Ni) were investigated by electron spin resonance. In these works,
the ESR spectra associated with axial Ni$^{3+}$-$V_{\rm O}$ complexes were
systematically observed along with the ESR spectra which were attributed to
the ``cubic'' Ni$^{2+}$ and Ni$^{3+}$ centers.~\cite{PhysRev.186.361}
Unfortunately, no arguments on the basis of which the oxidation state of the
Ni impurity was identified were presented in Ref.~\onlinecite{PhysRev.186.361},
and a possible interpretation of the ESR spectra as the spectra of Ni$^{4+}$
was not discussed.

\section{Conclusions}

The study of Ni-doped SrTiO$_3$ using X-ray diffraction and XAFS spectroscopy
enabled to draw the following conclusions:

1. The preparation conditions of single-phase Ni-doped SrTiO$_3$ samples with
the impurity concentration up to at least 3\% have been found. When incorporating
the impurity into the $A$~sites, the NiTiO$_3$ second phase was precipitated.
Regardless of the preparation conditions, the SrTi$_{1-x}$Ni$_x$O$_3$ solid
solution and NiTiO$_3$ phase are the most stable phases which can coexist.

2. EXAFS analysis showed that in the single-phase SrTi$_{0.97}$Ni$_{0.03}$O$_3$
sample, the Ni atoms substitute for the Ti atoms and are on-center. The
distortion of the oxygen octahedra, which could occur because of the Ni
off-centering or the existence of the oxygen vacancies in the Ni environment,
was not detected. The reduced value of the coordination number for the first
shell was explained by the presence of a small amount of Ni$^{3+}$-$V_{\rm O}$
complexes and the NiTiO$_3$ phase at the grain boundaries.

3. XANES analysis showed that in NiTiO$_3$ the oxidation state of Ni is 2+ and
in the SrTi$_{1-x}$Ni$_x$O$_3$ solid solution it is close to 4+.

4. The strongest light absorption in doped samples is associated with the
presence of tetravalent nickel in the SrTi$_{1-x}$Ni$_x$O$_3$ solid solution.
This doping seems to be the most promising for solar energy converters that
exploit the bulk photovoltaic effect.

In the future, we plan to perform similar experiments on the ferroelectric
BaTiO$_3$ doped with $3d$ elements in order to test the feasibility of obtaining
the polar material with similar optical and physical properties, in which the
bulk photovoltaic effect can be observed for photons in the whole spectrum
of solar radiation.

\begin{acknowledgments}
This work was supported by Russian Foundation for Basic Research (Grant No.
13-02-00724). I.A.S. and A.I.L. are grateful to Russian-German laboratory for
hospitality and financial support during their stay at BESSY.
\end{acknowledgments}

%\bibliography{all}

%\providecommand{\BIBYu}{Yu}
%

\end{document}